\documentclass[aps,prb,twocolumn,showpacs,superscriptaddress]{revtex4} 

\usepackage{amsfonts}
\usepackage{amsmath}
\usepackage{amssymb}
\usepackage{graphicx}
\usepackage{bm}

\begin{document}


\title{Resistivity Anisotropy of $AE$Fe$_2$As$_2$ ($AE$ =Ca, Sr, Ba): direct versus Montgomery technique measurements. }


\author{M.~A.~Tanatar}
\email[Corresponding author: ]{tanatar@ameslab.gov}
\affiliation{Ames Laboratory, Ames, Iowa 50011, USA}

\author{N.~Ni}

\affiliation{Ames Laboratory, Ames, Iowa 50011, USA}
\affiliation{Department of Physics and Astronomy, Iowa State University, Ames, Iowa 50011, USA }

\author{G.~D.~Samolyuk}
\altaffiliation[Permanent address:]{ Materials Science and Technology Division, Oak Ridge National Laboratory, Oak Ridge, TN 37831, USA.}
\affiliation{Ames Laboratory, Ames, Iowa 50011, USA}

\author{S.~L.~Bud'ko}
\affiliation{Ames Laboratory, Ames, Iowa 50011, USA}
\affiliation{Department of Physics and Astronomy, Iowa State University, Ames, Iowa 50011, USA }

\author{P.~C.~Canfield}
\affiliation{Ames Laboratory, Ames, Iowa 50011, USA}
\affiliation{Department of Physics and Astronomy, Iowa State University, Ames, Iowa 50011, USA }

\author{R.~Prozorov}
\email[Corresponding author: ]{prozorov@ameslab.gov}
\affiliation{Ames Laboratory, Ames, Iowa 50011, USA}
\affiliation{Department of Physics and Astronomy, Iowa State University, Ames, Iowa 50011, USA }

\date{\today}


\begin{abstract}

The anisotropy of electrical resistivity was measured in parent compounds of the iron-arsenic high temperature superconductors, $AE$Fe$_2$As$_2$ with Alkali Earth elements $AE$ =Ca, Sr, Ba. Measurements were performed using both the Montgomery technique and direct resistivity measurements on samples cut along principal crystallographic directions. The anisotropy ratio $\gamma_{\rho} = \rho_c /\rho_a$ is well below 10 for all compounds in the whole temperature range studied (4 to 300~K), in notable contrast to previous reports. The anisotropy at room temperature increases from about 2 in Ca, to about 4 in Sr and Ba. In all compounds the resistivity ratio decreases on cooling through the structural/antiferromagnetic transition temperature $T_{SM}$, with the change mainly coming from stronger variation in $\rho_a$ as compared with $\rho_c$. This suggests that the transitions stronger affect the two-dimensional parts of the Fermi surface. We compare our experimental observations with band structure calculations, and find similar trend in the evolution of anisotropy with the size of $AE$ ion. Our results show that the electronic structure of the iron pnictides has large contribution from three-dimensional areas of the Fermi surface. 
\end{abstract}

\pacs{74.70.Dd,72.15.-v,74.25.Jb}




\maketitle



\section{Introduction}

Following the original ideas by Little \cite{Little} and Ginzburg \cite{Ginzburg}, it was generally accepted that low dimensionality of the electronic spectrum is an important pre-requisite for finding superconductivity with high transition temperatures. These ideas fueled study of superconductivity in chained and layered materials \cite{problemhighTc}, including transition metal chalcogenides \cite{chalcogenides}, organics \cite{Ishiguro}, and most recently cuprates \cite{Bednorz}, MgB$_2$ \cite{MgB2PCC} and Sr$_2$RuO$_4$ \cite{Maeno}. Discovery of the iron-arsenide superconductors \cite{Hosono,Rotter}, characterized by layered structure of Fe-As layeres sandwiched between the layers of different chemical composition, seem to suggest that electronic structure of these materials may be two-dimensional as well. 
The $3d$ electronic orbitals of iron make the main contribution to the electronic bands close to the Fermi energy in iron-arsenides, and thus high anisotropy is naturally expected. This was suggested by early band structure calculations \cite{iron_BandS}, and seemed to find support in high ratio of electrical resistivity when measured for current flowing perpendicular to the Fe-As plane ($\rho_c$) and along it ($\rho_a$), $\gamma_{\rho} \equiv \rho_c/\rho_a  \sim $100, as reported for the non-superconducting parent compounds BaFe$_2$As$_2$ \cite{BaFeresaniz} and SrFe$_2$As$_2$ \cite{SrFeresaniz}, as well as for superconducting Co-doped BaFe$_2$As$_2$. \cite{Coaniz} 
Following this suggestion, formation of an antiferromagnetic state in the parent 
compounds below a temperature $T_{SM}$ of simultaneous structural/magnetic transition (in the range from 137~K in $AE$ =Ba to 210~K in $AE$ =Sr) was assigned to formation of spin density wave gapping part of the Fermi surface due to nesting.

Recently, however, we found \cite{anisotropy} notably smaller anisotropy of the electrical resistivity, $\gamma _{\rho}$, of the superconducting upper critical fields $\gamma_H \equiv \frac{H_{c2ab}}{H_{c2c}}$,\cite{NiNiCo} of the London penetration depth $\gamma_{\lambda} \equiv \frac{\lambda_{c}}{\lambda_{ab}}$ and of the superconducting critical current $\gamma_j \equiv \frac {j_{ca}}{j_{cc}}$, in the optimally Co-doped Ba(Fe$_{0.926}$Co$_{0.074}$)$_2$As$_2$, $T_c \approx 23$~K. Moreover, the anisotropies had values very close to the theoretically expected relations between these quantities, $\gamma_H =\gamma_{\lambda}=\sqrt{\gamma_{\rho}}$.\cite{anisotropy}

Since anisotropy is an important parameter, both for the mechanism of the magnetic state formation in the parent compounds and for superconductivity in the doped iron arsenides, we have undertaken comprehensive characterization of the resistivity anisotropy in the parent compounds of $AE$Fe$_2$As$_2$. Our main finding is that anisotropy is generally very low, inconsistent with two-dimensional models of the electronic structure.

\section{Experimental}

Single crystals of BaFe$_2$As$_2$ and of SrFe$_2$As$_2$ were grown from FeAs flux from a starting load of metallic Ba(Sr) and FeAs, as described in detail elsewhere. \cite{NiNiCo} Crystals were thick platelets with sizes as big as 12$\times$8$\times$1 mm$^3$ and large faces corresponding to the tetragonal (001) plane. Single crystals of CaFe$_2$As$_2$ were grown from Sn flux, as described by Ni Ni {\it et al.} \cite{Cagrowth}. The crystal quality of all samples was confirmed with X-ray Laue measurements on single crystals, which found resolution limited narrow peaks, see Refs.~[\onlinecite{NiNiCo,Cagrowth}] for details. No traces of either FeAs flux (by single crystal x-ray) or Sn flux (with wavelength dispersive electron probe microanalysis) were found. 

In our study of resistivity anisotropy in optimally doped superconducting Ba(Fe$_{1-x}$Co$_x$)$_2$As$_2$ with $x$=0.074 we have found that exfoliation of samples dramatically alters the sample resistivity, especially for measurements in configurations with current along tetragonal $c$-axis.\cite{anisotropy} Due to the softness of the materials, their cutting and shaping into transport samples inevitably introduces cracks, which affect the effective geometric factors of the sample. A strong tendency to exfoliate prevents the cutting of samples with $c \gg a$. Partial cleaving by exfoliation is one of the most likely reasons for the unusually high anisotropy, as found in previous studies. \cite{BaFeresaniz,SrFeresaniz,Coaniz} 

Samples for electrical resistivity measurements with current flow along the [100] $a$-axis in the tetragonal plane ($\rho _a$) were cut into bars of $(2-3) \times (0.1-0.2) \times (0.1-0.2)$ mm$^3$ ($a\times b\times c$). Samples for electrical resistivity measurements with current flow along the tetragonal $c$ axis ($\rho_c$) were cut into (0.3-0.7)$\times$(0.3-0.7)$\times$(0.1-0.5)mm$^3$ ($a\times b\times c$) bars. All sample dimensions were measured with an accuracy of about 10\%. Contacts to the samples were made by attaching silver wires with a silver alloy, resulting in an ultra low contact resistance (less than 100 $\mu \Omega$). Measurements of $\rho_a$ were made in both standard 4-probe and 2-probe configurations and gave identical results, see Ref.~[\onlinecite{vortex}]. Measurements of $\rho_c$ were made in the two-probe sample configuration. Contacts were covering the whole $ab$ plane area of the $c$-axis samples. A four-probe scheme was used to measure the resistance down to the contact to the sample, i.e. the sum of the actual sample resistance $R_s$ and contact resistance $R_c$ was measured. Taking into account that $R_s \gg R_c$, this represents a minor correction of the order of 1 to 5\%. This can be directly seen for superconducting samples \cite{anisotropy} at temperatures $T<T_c$, where $R_s =$0 and the measured resistance represents $R_c$.

Samples for Montgomery technique measurements of the resistivity anisotropy ratio, $\gamma_{\rho}$, were typically of the same size as samples for $\rho_c$ measurements. They had a ratio of sample dimensions along $a$, $l_a$, and along $c$, $l_c$, between 2 and 3. For resistivity anisotropy measurements in the $ac$ plane, contacts were made over whole lengths of the four sample edges along $b$-direction, with length $l_b$. A projection of the contacts on the $ac$ plane is schematically shown in inset in Fig.~\ref{Ca_Montg} (below). Four-probe resistivity measurements were made by sending current $I_1$ along one side of the sample (between contacts 1-4 for current along $a$) and measuring voltage $V_1$ on the opposite side (between contacts 2-3). Thus we determined the resistance $R_1=V_1/I_1$. In a next step the direction of the current was rotated by 90 degrees, with $I_2$ flowing along $c$-axis between contacts 1-2 and voltage $V_2$ measured between contacts 3 and 4. The ratio was used to define $R_2=V_2/I_2$. 
The ratio of the measured resistances, $R_1/R_2$ was used to determine the ratio of effective sample dimensions, $l'_a/l'_c$, using calculations of Ref.~[\onlinecite{Montgomery2}]. A comparison of the actual ($l_a/l_c$) and the effective ($l'_a/l'_c$) sample dimensions was used to determine the resistivity anisotropy as $(\rho_2/\rho_1)^{1/2}=(l'_c/l'_a)/(l_c/l_a)$.\cite{Montgomery1} Since the whole idea of the Montgomery technique is based on a homogeneous current distribution in the sample, the structural integrity of the sample, as well as the lack of inclusions of foreign materials (flux and solder) and voids play crucial role in measurements of this type. 

Because the contacts to the samples were extended over the whole length along $l_b$ we used a thin slab approximation \cite{Montgomery1} in the data analysis. During analysis we assumed the precise position of the contacts at the corners of the sample and neglected their size in the basal $ac$ plane of the rectangular prism. Since (1) the actual size of the contacts is 10 to 20\% of the sample dimensions and is not negligible as compared to either $l_a$ or $l_c$; (2) the contact positions and shape can deviate from ideal; (3) the sample section in the $ac$ plane is often not ideal and deviates from the assumed perfect rectangular, these factors bring sizable errors into the estimated anisotropy. To make our best effort, we were reproducing the results of Montgomery resistivity anisotropy measurements on several samples of each compound. We estimate the systematic error of these measurements as on the order of $\pm$ 50\% for the anisotropy ratio. 

Band structure calculations were performed using the full potential linearized augmented plane wave (FLAPW) approach \cite{FLAPW} and the local density approximation (LDA).\cite{LDA} The mesh of 31x31x31 $\vec k$-points was used for the Brillouin zone integration. We have used experimental lattice constants for the BaFe$_2$As$_2$ Ref.[\onlinecite{Ba_latticeconstants}], SrFe$_2$As$_2$ Ref.[\onlinecite{Sr_latticeconstants}] and CaFe$_2$As$_2$ Ref.[\onlinecite{Ca_latticeconstants}]. The Fermi velocities were calculated using the Bolz-Trap package.\cite{BolzTrap}

\section{Results}

\subsection{CaFe$_2$As$_2$ }

\begin{figure}
	
		\includegraphics[width=8.5cm]{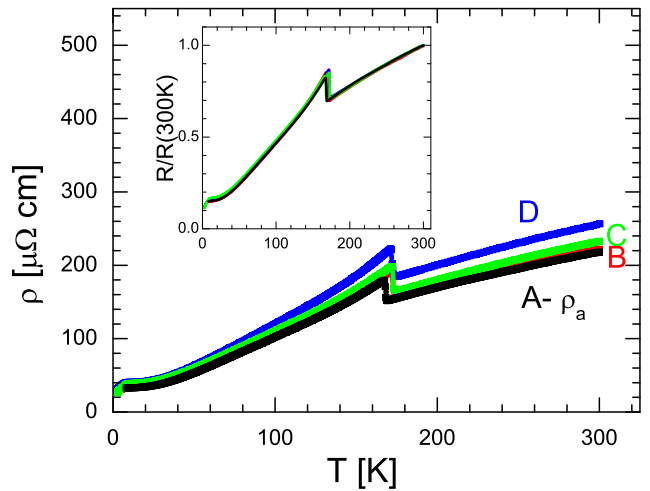}
	\caption{(Color online) Temperature dependence of $\rho_a $ measured on 4 different samples of CeFe$_2$As$_2$ (main panel). Inset shows the same data normalized to a value at 300~K. Samples are labeled A to D in line with increase of the room temperature resistivity value. }
	\label{Ca_rhoa}
\end{figure}

\begin{figure}
	
		\includegraphics[width=8.5cm]{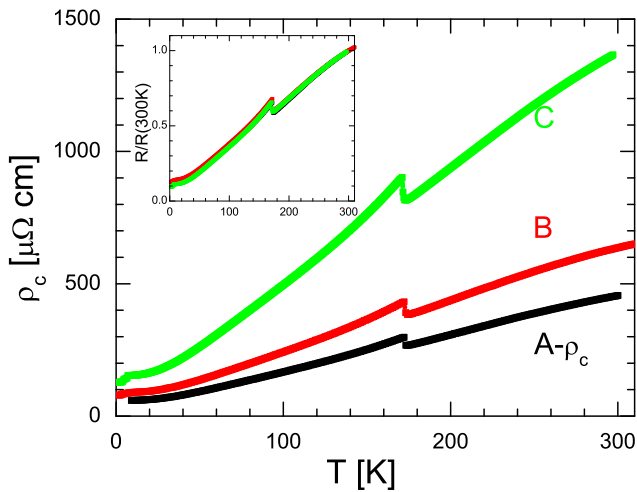}
	\caption{(Color online) Temperature dependence of $\rho_c $ measured on 3 samples of CaFe$_2$As$_2$ (main panel). Inset shows the same data normalized by the value of $rho_c$ at 300~K. The curves are labeled alphabetically in line with in room temperature resistivity increase. }
	\label{Ca_rhoc}
\end{figure}

\begin{figure}
	
		\includegraphics[width=8.5cm]{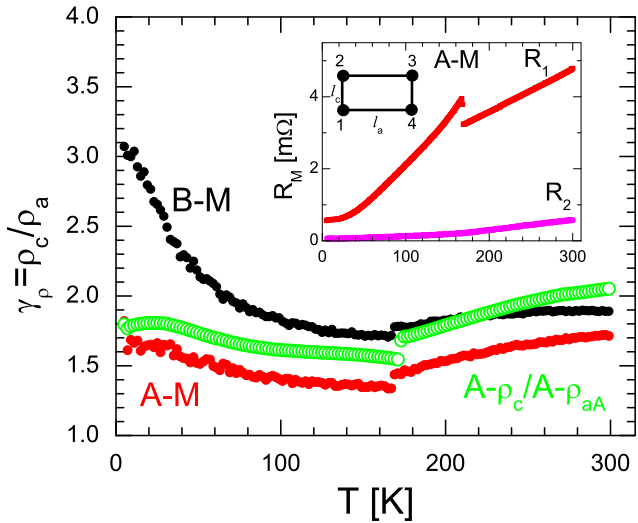}
	\caption{(Color online) Temperature dependence of the ratio of resistivities $\gamma_{\rho} \equiv \rho_c / \rho_a$ as determined from the Montgomery technique (solid symbols) and from the ratio of independently measured $\rho_a$ and $\rho_c$ for samples with the lowest resistivities, A-$\rho_c$ and A-$\rho_a$ (open symbols). Inset shows raw $R_1$ and $R_2$ data for sample A-M, and schematics of contact configuration during anisotropy measurements using Montgomery technique. Sample has rectangular cross-section in the $ac$ plane with dimensions $l_1$ (along $a$) and $l_2$ (along $c$). Contacts are located at the corners of the rectangle and run in the direction perpendicular to the plane of the figure along $l_3$ (parallel $b$). Two sets of four-probe resistivity measurements are performed: in the first run current $I_1$ is flowing between 1 and 4 along $l_1$, voltage $V_1$ is measured between contacts 2 and 3, their ratio determines resistance $R_1 =V_1/I_1$; in the second run direction of the current $I_2$ is along $l_2$ between contacts 1 and 2, voltage $V_2$ is measured between contacts 3 and 4, and $R_2= V_2/I_2$. The anisotropy $\gamma_{\rho}$ is calculated using $R_2/R_1$ and $l_2/l_1$ ratios as discussed in Refs.~[\onlinecite{Montgomery1,Montgomery2}]. 
}
\label{Ca_Montg}
\end{figure}

In Fig.~\ref{Ca_rhoa} and Fig.~\ref{Ca_rhoc} we show the temperature dependence of the electrical resistivities $\rho_a$ and $\rho_c$ (main panels in the figures) and the same data normalized by room temperature values (insets). The data with current along the plane was taken for 4 samples, the data for inter-plane current for 3 samples. The shape of the temperature dependent resistivity is very well reproduced between the samples of the same sort. Moreover, the value of the in-plane resistivity is very well reproduced as well, with the value at room temperature 232 $\pm$ 15 $\mu \Omega cm$ staying within the uncertainty of the geometric factor determination. The value of the inter-plane resistivity is more spread, from 456 $\mu \Omega cm$ for sample A to 1365 $\mu \Omega cm$ for sample C. Average over 3 samples gives a value of 800$\pm$400 $\mu \Omega cm$. As we have shown in our previous report on the superconducting samples, \cite{anisotropy} the lowest values are the most reliable, since the effect of exfoliations tends to increase $\rho_c$. This finds direct support in case of parent compounds, as shown by the comparison with Montgomery technique measurements, Fig.~\ref{Ca_Montg}. The inset shows the temperature dependence of the raw data, $R_1$ and $R_2$, for sample A with Montgomery contact configuration, main panel shows comparison of the $\gamma_{\rho}=\rho_c/\rho_a$ ratio as determined for 2 samples in Montgomery configuration and from direct $\rho_a$ and $\rho_c$ measurements for samples A (lowest resistivity) with in-plane and inter-plane currents. There is semi-quantitative agreement between $\gamma_{\rho}$ at room temperature determined in these two very different ways, with $\gamma_{\rho}$=1.7 for sample A-M, 1.9 for sample B-M and 2.1 for the comparison of direct resistivity measurements using the data for the lowest resistivity for $\rho_c$. Comparison of the average $\rho_a$ and $\rho_c$ values gives a ratio of about 4, notably larger than found in Montgomery technique measurements. 

It should be pointed out that good correspondence between the temperature dependent anisotropy ratio as determined from three different measurements is a very strong argument for the correct anisotropy determination. As can be seen from comparison of insets in Fig.~\ref{Ca_Montg} and Fig.~\ref{Ca_rhoa}, $R_1$ and $\rho_a$ reveal quite different temperature dependence, despite the fact that both are using in-plane current. This difference is even more striking when comparing measurements with inter-plane current flow, $R_2$ and $\rho_c$ (insets in Fig~\ref{Ca_Montg} and Fog.~\ref{Ca_rhoc}): while the former does not show even a trace of resistance increase at $T_{SM}$, the latter reveals it clearly. 

The ratio, as determined from independent measurements of $\rho_a$ and $\rho_c$, can be affected by the difference in sample quality. Sample resistivity can be written as $\rho= \rho_{0}+\rho_{in}$ and the residual resistivity $\rho_0$ can vary from sample to sample. This difference in $\rho_0$ is actually revealed by the comparison of the normalized curves in the insets of Fig.~\ref{Ca_rhoa} and Fig.~\ref{Ca_rhoc}. Therefore, the behavior of the temperature-dependent anisotropy reflects the intrinsic properties the best at high temperatures. Montgomery technique measurements are performed on one sample and therefore they are affected to a lesser extent by this difference. 
Based on these considerations we conclude that measurements on sample A-M, characterized by the smallest difference of $R_1$ and $R_2$, are most representative for intrinsic temperature-dependent anisotropy.

Of note is the temperature dependence of $\rho_a$ and $\rho_c$. Previous measurements found identical $\rho_a(T)$ curve with notable increase of resistivity below the structural/antiferromagnetic transition at 173~K. \cite{Cagrowth} This decrease reflects presumably partial loss of the density of states. The transition shows a hysteresis of about 2~K. The temperature dependence of $\rho_c$ is similar, however, here the increase of resistivity below the transition is notably smaller. As a result, the resistivity anisotropy decreases below the transition. On further cooling, though, the anisotropy continues to increase but only slightly. 


\subsection{SrFe$_2$As$_2$ }

\begin{figure}
	
		\includegraphics[width=8.5cm]{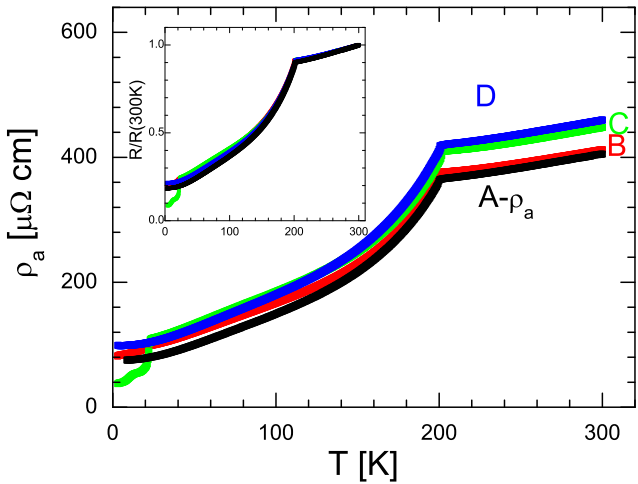}
	\caption{(Color online) Temperature dependence of $\rho_a $ measured on 4 different samples of SrFe$_2$As$_2$ (main panel). Inset shows the same data normalized to a value at 300~K. Samples are labeled A to D in line with increase of the room temperature resistivity value. }
	\label{Sr_rhoa}
\end{figure}

\begin{figure}
	
		\includegraphics[width=8.5cm]{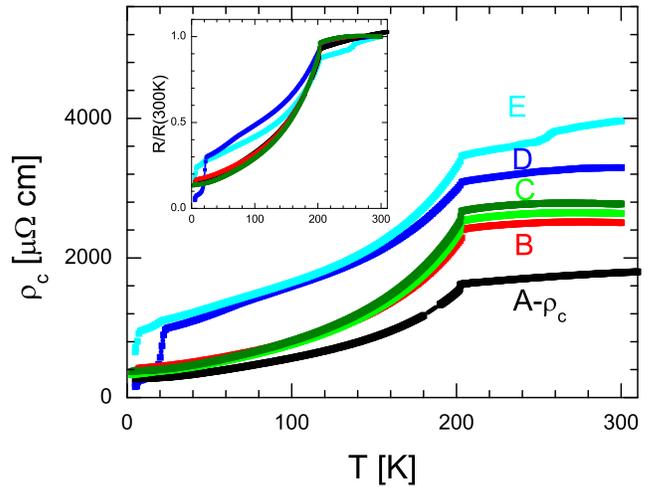}
	\caption{(Color online) Temperature dependence of $\rho_c $ measured on 5 samples of SrFe$_2$As$_2$ (main panel). Sample C was measured twice to check the effect of thermal cycling on resistivity value, the lower curve represents initial run. Inset shows the same data normalized by the value of $rho_c$ at 300~K. The curves are labeled alphabetically in line with room temperature resistivity increase. }
	\label{Sr_rhoc}
\end{figure}

In Fig.~\ref{Sr_rhoa} and Fig.~\ref{Sr_rhoc} we show the temperature dependence of the electrical resistivities $\rho _a$ and $\rho_c$ (main panels in the figures) and the same data normalized by the room temperature values (insets in the Figures) for SrFe$_2$As$_2$. The data with current along the plane was taken for 4 samples, the data for inter-plane current for 5 samples.

The value of the in-plane resistivity is well reproduced within error bars of geometric factor determination, at room temperature it is 430 $\pm$ 24 $\mu \Omega cm$. The shape of the temperature dependent resistivity, $\rho_a (T)$,  is well reproduced as well, see inset in Fig.~\ref{Sr_rhoa}. On cooling $\rho_a$ gradually decreases, the rate of decrease increases below $T_{SM}$ =205~K. A close inspection of the data reveals a tiny increase in the very vicinity of the transition (hard to see in the Fig.~\ref{Sr_rhoa}). 

The value of the inter-plane resistivity is more spread, from 1790 $\mu \Omega cm$ for sample A to 3960 $\mu \Omega cm$ for sample E. Note, that samples with higher resistivity show notably different temperature dependence (inset in Fig.~\ref{Sr_rhoc}). In addition, they clearly show traces of superconductivity. Since superconductivity in pure SrFe$_2$As$_2$ is induced by residual strain, \cite{Paglione} as well as by hydrostatic pressure \cite{Sr-SCpressure}, this allows us to link the abnormally high resistivity values with the presence of deformed areas in the samples. This agrees well with the conclusion of our previous study, linking high resistivity values with partial exfoliation,\cite{anisotropy} in which case strained regions are easily formed, as well as with small volume fraction of superconducting inclusions found in magnetization studies \cite{Paglione}. 
The temperature dependence of inter-plane resistivity, $\rho_c(T)$, shows the same features as $\rho_a(T)$, however, a flattening above $T_{SM}$ and the decrease below $T_{SM}$ are more pronounced in $\rho_c$. 

\begin{figure}
	
		\includegraphics[width=8.5cm]{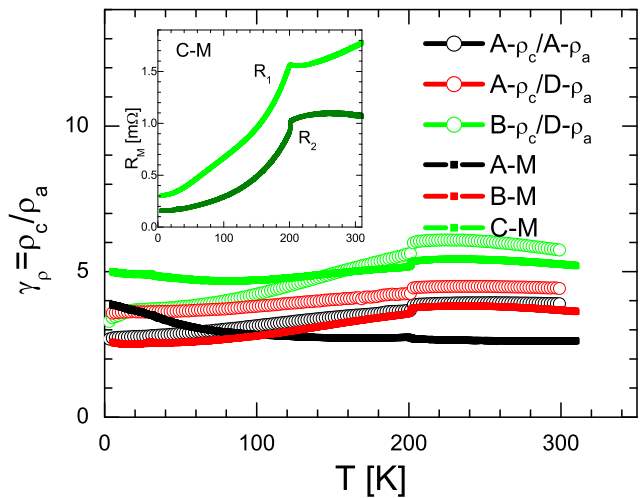}
	\caption{(Color online) Temperature dependence of the ratio of resistivities $\gamma_{\rho} \equiv \rho_c / \rho_a$ as determined from the Montgomery technique (solid symbols) and from the ratio of independently measured $\rho_a$ and $\rho_c$ for several different samples (open symbols). Inset shows raw $R_1$ and $R_2$ data for sample C-M measured in Montgomery configuration. 
}
\label{Sr_Montg}
\end{figure}

The $\gamma_{\rho}$ anisotropy ratio in SrFe$_2$As$_2$, as determined from direct measurements and from Montgomery technique measurements, are plotted in Fig.~\ref{Sr_Montg}. Larger open symbols show the temperature dependence of the $\gamma_{\rho}$ ratio of resistivities for selected samples: samples with lowest measured resistivities at room temperature, A-$\rho_c$/A-$\rho_a$, sample with lowest $\rho_c$ with sample with highest $\rho_a$, A-$\rho_c$/D-$\rho_a$, as well as ratio of second highest resistivity sample B for inter-plane resistivity and highest resistivity sample D for in-plane resistivity, B-$\rho_c$/D-$\rho_a$. Small solid symbols show $\gamma_{\rho}$ as calculated following Montgomery technique procedure based on the $R_1$ and $R_2$ raw data (shown for sample C in inset of Fig.~\ref{Sr_Montg}). There is very good general agreement between the two measurements. Note, that the ratio of the lowest $\rho_c$ and $\rho_a$ reproduces well the temperature dependence as well as magnitude of the $\gamma_{\rho} $ for two samples measured in Montgomery technique. The most characteristic features of this behavior are slight increase of the anisotropy on approaching the structural/magnetic transition from above, sharp drop of the anisotropy at the transition and small $\gamma_{\rho}$ decrease on further cooling. 

Of note, neither the value of the anisotropy nor its temperature dependence of inter-plane resistivity, $\rho_c(T)$, are consistent with findings of the previous study.\cite{SrFeresaniz} The difference of magnitude is dramatic, with anisotropy in the range 3 to 5 in our case as compared with 130 in the previous study. Since no detail of the measurement procedure or its reproducibility, sample to sample, is given in Ref.~[\onlinecite{SrFeresaniz}], we can not comment on the nature of the difference. We speculate that heavy contamination of the measured $\rho_c$ data with $\rho_a$ and strong deviations of the current path from the projected one must be the reason for the incorrect anisotropy determination in a previous study.


\subsection{BaFe$_2$As$_2$ }

\begin{figure}
	
		\includegraphics[width=8.5cm]{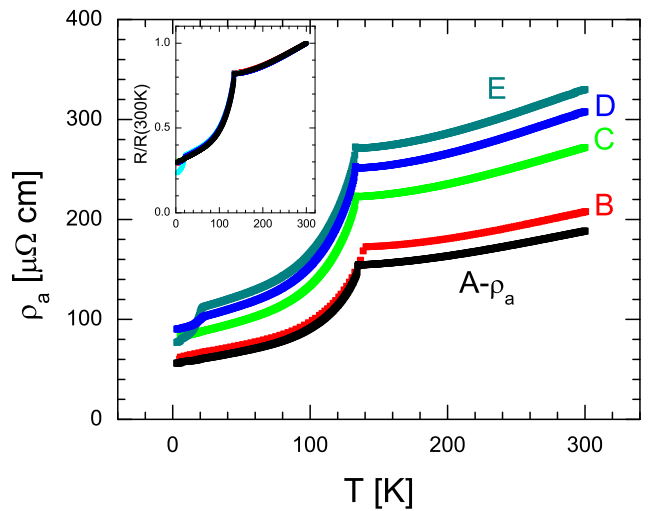}
	\caption{(Color online) Temperature dependence of $\rho_a $ measured on 5 different samples of BaFe$_2$As$_2$ (main panel). Inset shows the same data normalized to the resistivity values at 300~K. Samples are labeled A to E in line with increase of the room temperature resistivity. }
	\label{Ba_rhoa}
\end{figure}

\begin{figure}
	
		\includegraphics[width=8.5cm]{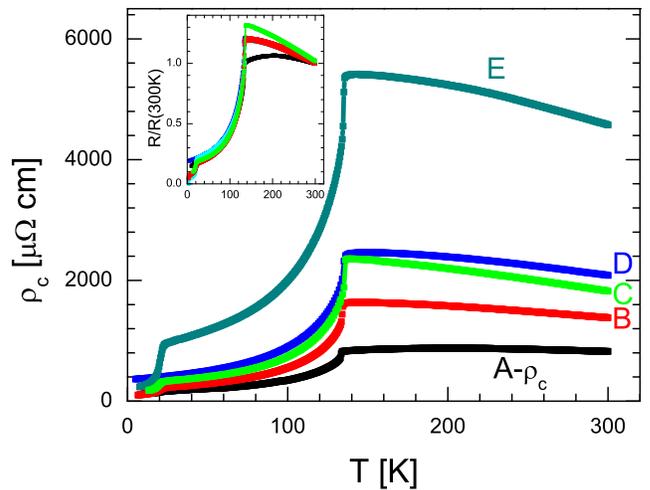}
	\caption{(Color online) Temperature dependence of $\rho_c $ measured on 5 samples of BaFe$_2$As$_2$ (main panel). Inset shows the same data normalized by the values of $rho_c$ at 300~K. The curves are labeled alphabetically in line with room temperature resistivity increase. }
	\label{Ba_rhoc}
\end{figure}

In Fig.~\ref{Ba_rhoa} and Fig.~\ref{Ba_rhoc} we show the temperature dependence of the electrical resistivities $\rho _a$ and $\rho_c$ (main panels in the figures) and the same data normalized by room temperature values (insets in the figures) for BaFe$_2$As$_2$. The data for both current directions were taken for sets of 5 samples each. 

The value of the in-plane resistivity in BaFe$_2$As$_2$ is the most scattered among the three compounds. The spread notably exceeds the error bars of geometric factor determination, giving the value at room temperature of 260 $\pm$ 55 $\mu \Omega cm$. The shape of the resistivity temperature dependence $\rho_a (T)$ is reproduced well, with the only exception for the presence of partial superconducting transition in the samples D and E with the highest resistivity. This again suggests that these samples have strained regions, relating them with partial cracks. On cooling $\rho_a$ gradually decreases, with the decrease rate being enhanced below $T_{SM}$=137~K.

The value of the inter-plane resistivity is even more spread, from 910 $\mu \Omega cm$ for sample A to 4540 $\mu \Omega cm$ for sample E. Note, that similar to SrFe$_2$As$_2$, the samples with higher resistivity show more pronounced superconducting feature, suggesting the existence of areas with high internal pressure. The value over 5 samples averages to 1760$\pm$1310 $\mu \Omega cm$; when excluding from consideration the outstanding sample E, averaging gives 1490$\pm$390 $\mu \Omega cm$. The shape of the temperature dependent resistivity (inset in Fig.~\ref{Sr_rhoc}), is not reproduced as well as that for Ca and Sr compounds. Its pronounced feature is the increase of $\rho_c$ on cooling down from room temperature. For four samples the increase continues all the way down to $T_{SM}$, followed by a decrease typical of a metal below. In one of the samples, (A-$\rho_c (T)$ curve in Fig.~\ref{Ba_rhoc}), the $\rho_c(T)$ shows a broad maximum at around 200~K and practically flattens below in the range down to $T_{SM}$.  

\begin{figure}
	
		\includegraphics[width=8.5cm]{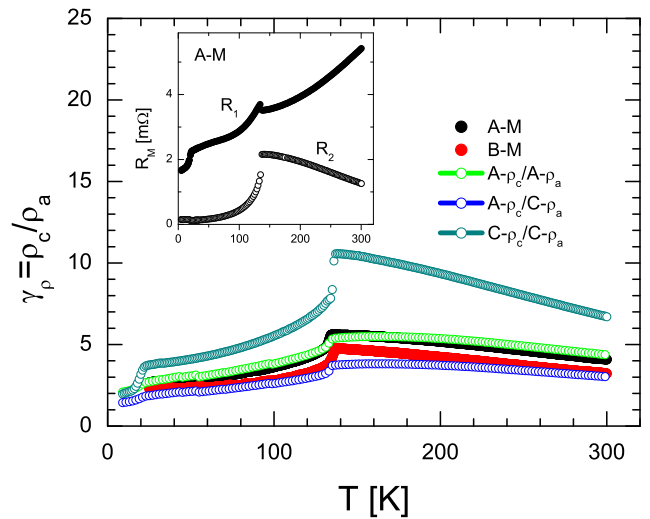}
	\caption{(Color online) Temperature dependence of the ratio of resistivities $\gamma_{\rho} \equiv \rho_c / \rho_a$ as determined from the Montgomery technique (solid symbols) and from the ratio of independently measured $\rho_a$ and $\rho_c$ for some representative samples (open symbols). Inset shows raw $R_1$ and $R_2$ data for sample C-M measured in Montgomery configuration. 
}
\label{Ba_Montg}
\end{figure}

The $\gamma_{\rho}$ anisotropy ratio in BaFe$_2$As$_2$, as determined from direct measurements and from Montgomery technique measurements, are plotted in Fig.~\ref{Ba_Montg}. The larger open symbols show the temperature dependence of the $\gamma_{\rho}$ ratio of resistivities for selected samples: samples with lowest measured resistivities at room temperature, A-$\rho_c$/A-$\rho_a$, sample with lowest $\rho_c$ with sample with medium $\rho_a$, A-$\rho_c$/C-$\rho_a$, as well as ratio of the medium resistivity sample C for inter-plane resistivity and medium resistivity sample C for in-plane resistivity, C-$\rho_c$/C-$\rho_a$. Small solid symbols show $\gamma_{\rho}$ calculated following Montgomery technique procedure, using $R_1$ and $R_2$ raw data shown for sample A-M in inset of Fig.~\ref{Ba_Montg}. The two ways of resistivity anisotropy determination agree in general, the agreement being the best when comparing the ratios obtained for lowest $\rho_c$. Anisotropy at room temperature is between 3 and 5, similar to SrFe$_2$As$_2$. The spread of the ratios is bigger, due to bigger scatter in the resistivity value both for $\rho_a$ and $\rho_c$.

The most characteristic features the temperature dependence of $\gamma_{\rho}$ in BaFe$_2$As$_2$ are increase in anisotropy on cooling down to $T_{SM}$, sharp drop at the transition, and slow decrease of the anisotropy below the transition.


\section{Discussion}

\subsection{Resistivity anisotropy versus size of $AE$ element}

\begin{figure}
	
		\includegraphics[width=8.5cm]{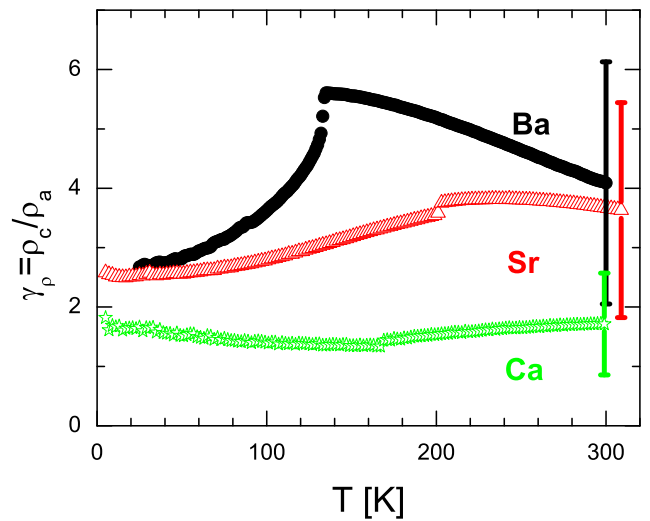}
	\caption{(Color online) Temperature dependence of the anisotropy ratio, $\gamma_{\rho}=\rho_c/\rho_a$ for CaFe$_2$As$_2$, SrFe$_2$As$_2$ and BaFe$_2$As$_2$. The error bars represent the evaluated systematic error of the anisotropy determination. 
}
\label{gamma_comparison}
\end{figure}

The comparison of the Montgomery and direct resistivity anisotropies for the three different $AE$Fe$_2$As$_2$ compounds suggests that the best agreement between the two types of anisotropy determinations is obtained when using samples with the lowest ratio of $R_1$ and $R_2$ resistances in the Montgomery measurements and the lowest resistivity value for $\rho_c$. This is consistent with the conclusion of our previous study \cite{anisotropy} on resistivity anisotropy in the superconducting BaFe$_2$As$_2$ doped with Co. We now turn to comparison of the data for different compounds using these criteria for selection. 

In Fig.~\ref{gamma_comparison} we compare the temperature-dependent anisotropy ratios for the three $AE$Fe$_2$As$_2$ compounds. Despite sizable systematic error, comparison reveals a clear trend in the evolution of $\gamma_{\rho}(T)$. First, at room temperature the ratio is the lowest in the Ca compound ($\sim$2), while the anisotropy ratios of the Sr and Ba compounds (about 4) are the same within error bars. Both these numbers are too low to be discussed in the two-dimensional Fermi surface scenario. 

The temperature-dependent anisotropy reveals systematic evolution with the size of the $AE$ atom as well. In the Ca compound $\gamma_{\rho}$ decreases on cooling from room temperature down to $T_{SM}$, shows a down-jump at $T_{SM}$, and slightly increases below. In the Sr compound the anisotropy very slightly increases down to $T_{SM}$, jumps down at the transition and gradually decreases on further cooling. In the Ba compound the anisotropy increases gradually all the way down to $T_{SM}$, below which the anisotropy goes down very rapidly. Even though the $\gamma_{\rho}$ ratios for the Sr and Ba compounds are very close both at room temperature and at low temperatures, the anisotropy of the Ba compound becomes more pronounced in the intermediate temperature range, especially above $T_{SM}$.  Thus we conclude that the anisotropy follows the ionic radius of alkali earth elements, similar to the predictions of the band structure calculations. 

\subsection{ Band structure}

\begin{figure}

\includegraphics[width=8.5cm]{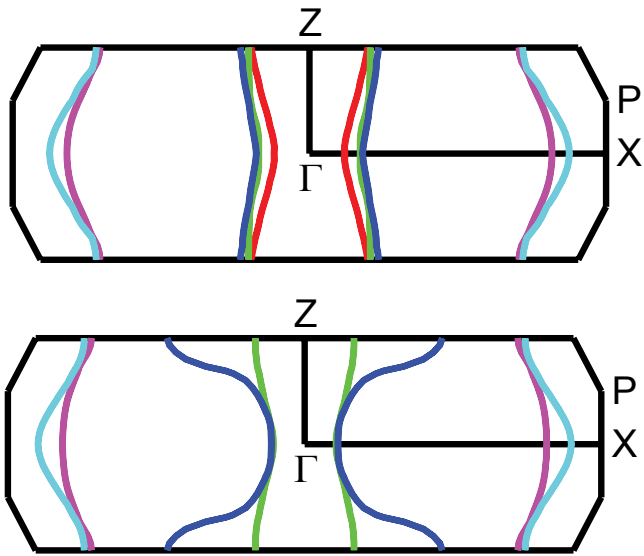}
	\caption{(Color online) Cross-section of the Fermi surface of BaFe$_2$As$_2$ by the $\Gamma X Z$ plane in the Brillouin zone. The position of the As atom in the lattice, $Z_{As}$, was assumed corresponding to experimentally measured \cite{Ba_latticeconstants} (top panel), or to the minimum of the total crystal energy (bottom panel). Variation of $Z_{As}$ affects the most the sheet of the Fermi surface surrounding the $\Gamma$ point in the Brillouin zone. 
}
\label{FS}
\end{figure}

\begin{table*} 
\caption{Calculated kinetic characteristics of $AE$Fe$_2$As$_2$ compounds in non-magnetic states. } 
\vspace{.1cm}  
\begin{tabular*}{17cm}{lccccccccc}\hline 
\vspace{.1cm}  
Compound & $~~~\sigma_{xx}/\tau$ ~~~& ~$~~~\sigma_{zz}/\tau$ ~~~&~~~ $\sigma_{xx}/\sigma_{zz}$ ~~& ~~~~~DOS ~~~~~& ~~~Vol. ~~~& ~~~$v_x$ ~~~& ~~~$v_z$ ~~~&~~$Z_{As}$ ~~& Reference\\  \hline\ 
\\  
Units & $10^{19}$/($\Omega $~m~s) & $ 10^{19}$/($\Omega $~m~s) &~&1/(eV f.u.) & \AA$^3$& $10^5~m/s$ & $10^5~m/s$ &~~~~&~~~~ \\  \hline\ 

\vspace{.1cm} 

CaFe$_2$As$_2$ collapsed & 15.1 & 23.6 &~0.64 & 4.5 & 566.02& 1.33 & 1.667 & 0.366 & [\onlinecite{Ca_latticeconstants}] \\  \hline\ 
\vspace{.1cm} 

CaFe$_2$As$_2$ & 18.7 & 12.5 & 1.5 & 6.8 & 602.45 & 1.24 & 1.01 & 0.372 & [\onlinecite{Ca_latticeconstants}] \\  \hline\ 
\vspace{.1cm} 

SrFe$_2$As$_2$ & 16.0 & 2.2 & 7.2 & 5.0 & 642.93 & 1.380 & 0.515 & 0.360 & [\onlinecite{Sr_latticeconstants}] \\  \hline\ 
\vspace{.1cm} 

BaFe$_2$As$_2$ & 14.0 & 1.1 & 12.1 & 4.7 & 690.31 & 1.380 & 0.398 & 0.355 & [\onlinecite{Ba_latticeconstants}] \\  \hline\ 
\vspace{.1cm} 

BaFe$_2$As$_2$ & 14.4 & 6.13 & 2.35 & 3.0 & 690.31 & 1.75 & 1.14 & 0.341 &calculated\\  \hline\ 
\vspace{.1cm}

\end{tabular*} 
\end{table*}

A key structural feature of the iron arsenide compounds, which makes a profound effect on their electronic structure, is the location of the As atoms above and below the layer of Fe atoms. The band dispersion along the tetragonal $c$-direction is mainly determined by the overlap of As orbitals. Thus the anisotropy of the electronic structure is extremely sensitive to the displacement coordinate $Z_{As}$ of As atom in the unit cell with respect to the Fe-layers.

Since the actual location of the As atoms can vary, in our calculations in the paramagnetic phase of BaFe$_2$As$_2$ we have used two positions. In the first one, the position of As atoms was determined from the calculated minima of the total energy. We refer to this as the calculated or relaxed position below. The position of the As atom, $Z_{As}$=0.341, as determined in our analysis, is very close to the positions found in previous calculations, $Z_{As}$=0.342,\cite{MP2} but is significantly lower than the experimental value of $Z_{As}$=0.355 at room temperature. \cite{Ba_latticeconstants} The Fermi velocities, calculated using the Bolz-Trap \cite{BolzTrap} package, are very sensitive to the As positions as well. A similar trend is found in magnetic properties.\cite{MP2,MP1} The downshift by 0.16 \AA\ increases both the band dispersion and the Fermi velocity along the $z$ direction. The calculations with relaxed $Z_{As}$  give $V_{Fa}^2/V_{Fc}^2$=3 for pure BaFe$_2$As$_2$, with the experimental $Z_{As}$ we come to much larger anisotropy of 12.1. These two anisotropy values should be compared with the anisotropy of about 4, as found for this compound in our experiment. The resistivity anisotropies using experimental $Z_{As}$ for all compounds follow ionic radius of Alkali Earth elements, similar to a trend found in the experiment. We summarize the calculated anisotropies in Table 1. 

The calculation shows that the anisotropy of the Fermi velocities, averaged over the Fermi surface sheets, varies a lot with $Z_{As}$. However, when looking at individual Fermi surface sheets, a notable difference in response to variation of $Z_{As}$ is found (Fig.~\ref{FS}). The shape of the central sheet of the Fermi surface, surrounding the $\Gamma$ point of the Brillouin zone, is most sensitive to the position of the arsenic atom, and for relaxed $Z_{As}$ it develops pronounced warping. Other sheets remain cylindrical, and therefore most correct description of the electronic structure should be as a combination of two- and three-dimensional portions. In this case contribution of the three-dimensional central sheet to $\rho_c$ would "shunt" very anisotropic contributions of the rest of the Fermi surface.

Obviously, our experimental values of the anisotropy are compatible with both anisotropic three-dimensional and multi-dimensional Fermi surface, including two- and three-dimensional sheets.Unfortunately, transport measurements alone cannot separate between these options. 

Interestingly enough, similar multi-dimensional Fermi surface, including two- dimensional (2D) and three-dimensional (3D) sheets is found in a number of superconductors with rather high transition temperatures. In MgB$_2$, the anisotropy of electrical resistivity, $\rho_c/\rho_{ab}$, is about 3.5,\cite{MgB2} while 2D and 3D sheets have about the same density of states. In borocarbides, the anisotropy of electrical resistivity is of the order of one \cite{borocarbidesAnisotropy}, while the Fermi surface is composed of both 3D and warped 2D sheets, the latter having well defined nesting areas. \cite{borocarbidesFS} In NbSe$_2$ the anisotropy ratio is about 30, \cite{NbSe2} due to a much larger contributions of the two-dimensional sections of the Fermi surface into transport. In the heavy fermion superconductor CeCoIn$_5$, the anisotropy ratio is small, \cite{CeCoIn5anisotropy} despite a presence of the two-dimensional sheet in the Fermi surface, \cite{CeCoIn52DFS} and temperature-dependent. This temperature dependent anisotropy reflects the difference in the temperature-dependent resistivity for two current flow directions \cite{Science} in CeCoIn$_5$ due to an anisotropy of magnetic scattering. 

The superconducting state in all these compounds is anomalous and frequently characterized by strong variation of the superconducting gap magnitude between different Fermi surface sheets. \cite{MgB2multiband,NbSe2Yokoya,NbSe2Boaknin,CeCoIn5Tanatar} It is therefore of great interest if the same is the case in the iron-arsenic superconductors. Strong temperature dependence of the anisotropy of London penetration depth in optimally doped Ba(Fe$_{1-x}$Co$_x$)$_2$As$_2$ ($T_c$~23~K), \cite{anisotropy} in (Ba,K)Fe$_2$As$_2$ ($T_c$~30~K), \cite{MartinBaK} and in NdFeAs(O,F), \cite{ProzorovPhysicaC} similar to the cases of MgB$_2$ (Ref[\onlinecite{MgB2penetration}]) and NbSe$_2$ (Ref[\onlinecite{NbSe2penetration}]), suggests that the situation in iron-arsenic family may be similar.

\subsection{Comparison of the resistivity temperature dependences}

\begin{figure}
		\includegraphics[width=8.5cm]{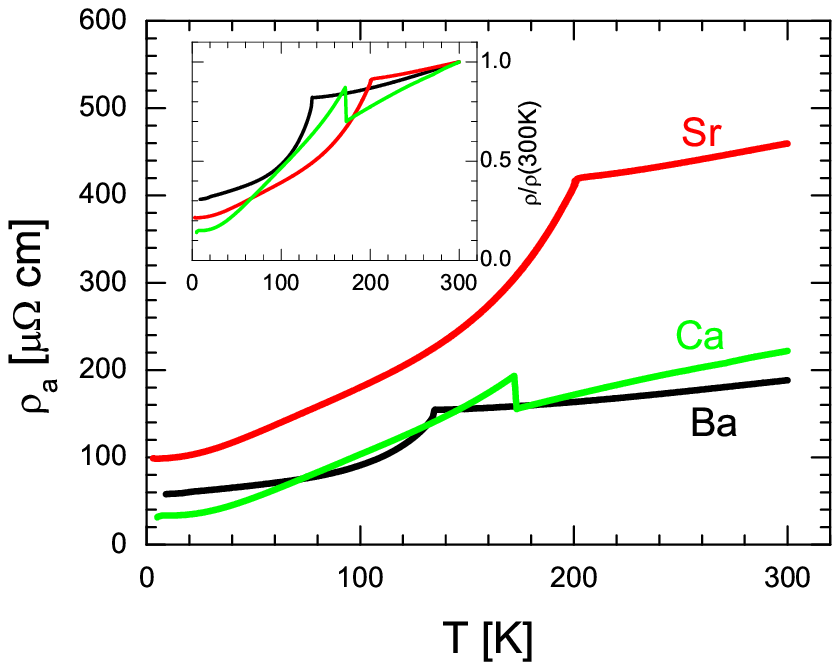}
		\includegraphics[width=8.5cm]{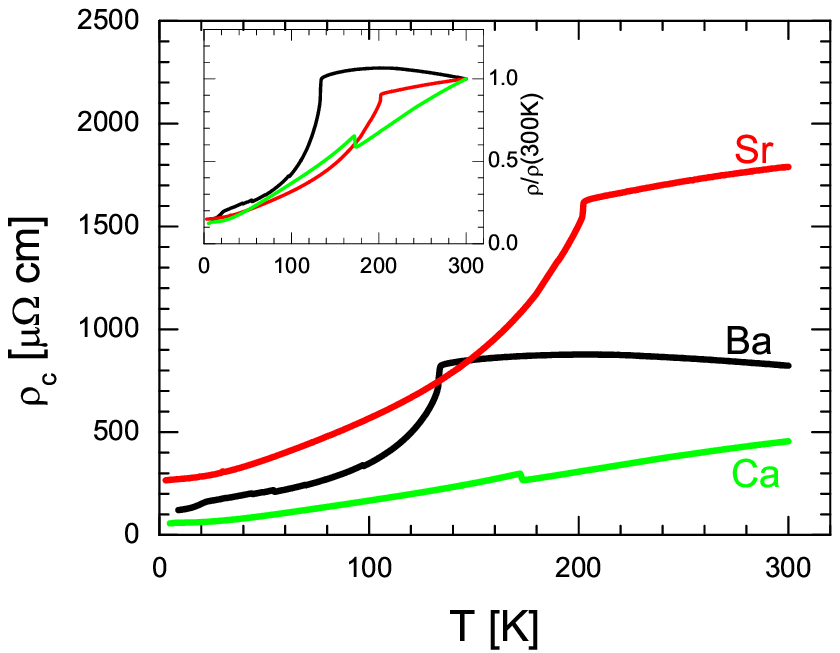}
	\caption{(Color online) Temperature dependence of the in-plane (top panel) and inter-plane (bottom panel) resistivity for CaFe$_2$As$_2$, SrFe$_2$As$_2$ and BaFe$_2$As$_2$. The curves were selected as corresponding to a minimum room-temperature resistivity value for each type of the curves. Insets show the same data normalized by the resistivity values at 300~K.
}
\label{rho_comparison}
\end{figure}

In the top panel of Fig.~\ref{rho_comparison} we show temperature-dependent in-plane resistivity of the three compounds, inset shows the same data normalized by the room-temperature values. Similar plots for the temperature-dependent inter-plane transport are shown in the bottom panel Fig.~\ref{rho_comparison}.

It is interesting to notice that, despite monotonic change of the anisotropy with the size of the $AE$ element, resistivity temperature dependences do not show systematic evolution. Moreover, the temperatures of the structural/antiferromagnetic transitions do not follow monotonic trend either, with $T_{SM}$ being maximum for Sr (205~K), followed by Ca (173~K) and Ba (137~K). The in-plane resistivity (top panel of Fig.~\ref{rho_comparison}) may be the only other quantity, which follows the same order as $T_{SM}$, if taking minimum resistivity values. (Strictly speaking, the resistivities of the Ca and Ba compounds at room temperature coincide within the error bars.) This similar trend in the value of the in-plane resistivity and $T_{SM}$ may be suggestive that the proximity to the transition is a factor important for the resistivity. It obviously suggests that pre-transition fluctuations play important role in the scattering already at room temperature. 

For all three compounds the anisotropy ratio decreases stepwise on passing through $T_{SM}$.  This fact is suggestive that the two-dimensional portions of the Fermi surface, contributing more to the $\rho_a$, are affected stronger by the transition. This is in line with the idea about possible role of Fermi surface nesting in the transition. 

On the other hand, the Ca compound, characterized by the lowest anisotropy, is the only one which clearly shows resistivity increase below the transition for both current flow directions. 
It also shows notably lower inter-plane resistivity, and clear metallic character of its resistance temperature dependence both above and below $T_{SM}$. This clearly suggest that the pre-transition fluctuations of the order parameter do not play as large role in the resistivity of the Ca compound as they do in the Ba and Sr compounds.

\section{Conclusions}

The parent compounds of iron arsenic superconductors reveal relatively small anisotropy of the electrical resistivity. The ratio of the inter-plane and in-plane resistivities stays well below 10 for all temperature range studied, and nowhere close to the reported values of about 100.\cite{BaFeresaniz, SrFeresaniz,Coaniz} The anisotropy increases in line with the ionic radius of $AE$ element, revealing the same trend as found in band structure calculations. We do not see any systematic evolution of the temperature dependent resistivity, either for the in-plane or for the inter-plane transport, with the size of the alkali earth element, same as the temperature of the structural/antiferromagnetic transition. On the other hand, there seems to be a correlation between the value of the in-plane resistivity at room temperature and $T_{SM}$. We speculate that pre-transition fluctuations play important role in scattering. 

\section{Acknowledgements}

We thank A. Kaminski and Y. Lee for discussions. M.A.T. acknowledges continuing cross-appointment with the Institute of Surface Chemistry, National Ukrainian Academy of Sciences. Work at the Ames Laboratory was supported by the Department of Energy-Basic Energy Sciences under Contract No. DE-AC02-07CH11358. R. P. acknowledges support from Alfred P. Sloan Foundation.


\end{document}